# Point vortices dynamics on a rotating sphere and modeling of global atmospheric vortices interaction

Igor I. Mokhov[1,2,3], S. G. Chefranov [*) 1,4], and A. G. Chefranov[5]

[1]*A.M. Obukhov Institute of Atmospheric Physics RAS, Moscow 119017, Russia*

[2]*Lomonosov Moscow State University, Moscow, Russia*

[3]*Moscow Institute for Physics and Technology, Dolgoprudny, Russia*

[4]*Physics Department, Technion-Israel Institute of Technology, Haifa 32000, Israel*

[5]*Eastern Mediterranean University, Famagusta 99628, North Cyprus*

## Abstract

It is shown that the hydrodynamics equations for a thin spherical liquid layer are satisfied by the stream function of a pair of antipodal vortices (APV), in contrast to the stream function of a single point vortex on a sphere with a background of a uniform opposite sign vorticity. A simple zero solution of the equation of the absolute vorticity conservation is used for bypassing well-known nonlinear problem of a point vortices interaction with regular vorticity field and an exact solution for APVs dynamics problem on a rotating sphere is obtained. Due to this a new stable stationary solution for the dynamics of APV is obtained, which can model the dynamics of the global vortex structures such as atmospheric centers of action.

. *) Corresponding Author: schefranov@mail.ru

## 1. INTRODUCTION

Study of the point vortices (PV) on a sphere is important for understanding the processes of the atmospheric and oceanic vortices evolution and modeling of them [1-10]. Conceptual models of point vortices are useful in fluid dynamics for identification and evaluation of physical mechanisms affecting the structure and interaction of atmospheric and oceanic vortices [11].

The key large-scale vortex structures in the atmosphere are atmospheric centers of action (ACA) [12–16]. Regional and large-scale atmospheric anomalies and their changes are related to the ACAs regimes. For example, atmospheric conditions in the Northern Hemisphere over Europe are related to quasi-steady Icelandic cyclone (IC) and Azores anticyclone (AA) over North Atlantic [14–16].

For modeling of dynamics of vortices like ACA dynamics the system of interacting antipodal point vortices (APV) is used [1,3,4,17]. Each APV is a combination of two point vortices located at diametrically conjugated points of the sphere and having equal magnitude, but opposite sign circulation. For each APV the Kelvin theorem on zero total vorticity on the sphere holds. The data available from observations of global vortices [18] also give grounds for using APVs in modeling of ACA-like dynamics.

Up to now absolute stationary solutions to the dynamics equations of APVs on a rotating sphere, which could simulate the stationary or quasi-stationary states like for the centers of action in the atmosphere are not known. The attempt to simulate of such modes using point-vortices on a sphere faces the problem of accounting for the influence of the sphere rotation on the dynamics of the PVs. The rotation of the Earth is an important factor determining the dynamics of global-scale vortices, for which these stationary states are observed precisely in a non-inertial rotating system of reference. Due to the non-uniform background vorticity coming from the Coriolis force the Hamiltonian system for APVs dynamics becomes infinite dimensional and is described by integral—differential equations [3]. Up to now it is known only relative stationary solutions for the APVs system, which cannot be used for modeling ACAs absolute stationary dynamical regimes of ACAs. Most of the works are based on approximate approaches for the sphere rotation [6,10,19,20]. Moreover the evolution of point-vortices nonlinear interacting with a rotating spherical liquid layer is considered there only in a fixed inertial system of reference. The corresponding approach, except the paper [21], used so far to describe point-vortices on a rotating sphere is purely kinematic and is not directly based on the hydrodynamic equations for a thin rotating spherical layer in the form of absolute vorticity conservation equation in a rotating non-inertial coordinate system [22]. On the other hand in [21] the theory of PV on the beta-plane is generalized to the case of a rotating sphere on the base of stationary solution to absolute vorticity conservation equation (see equation (1) in [21]). But in [21] also only relative steady-state of PVs system is obtained when this system is moving along a circle of the constant latitude with finite constant angular velocity $\omega \neq 0$. And thus it is impossible to use that solution for describing of absolute steady-state of ACA-like objects over oceans when condition $\omega = 0$ is required for realization of absolute steady-state regime in ACA-like dynamics.

Thus, the problem of modeling the observed absolute stationary state of global-scale cyclone-anticyclone pairs like ACAs over the oceans using the dynamics of point vortices on a rotating sphere remains unsolved.

In this paper, a solution is obtained for the dynamics of APV on a rotating sphere, for which there is a stable stationary mode that simulates exactly the absolute stationary ACA-like structure corresponding to the condition $\omega = 0$.

A simple non-trivial zero solution of the absolute vorticity conservation equation is used for the exact describing APV dynamics on the rotating sphere, when in contrast to [21] it is possible to realize condition $\omega = 0$. A zero value of absolute vorticity corresponds to a non-zero stream function on the sphere, which influenced on 2N- dimensional dynamical system of equations for the APVs on the rotating sphere [23–25]. Due to the zero value of absolute vorticity regular field the well-known [3,6,10,19–21] complex nonlinear problem of interaction between PVs and regular vorticity field is removed automatically and the APVs dynamics on a rotating sphere is exactly described. It allows obtaining a stationary solution to the dynamics equations of APVs on a rotating sphere, which can simulate the stationary ACA-like states.

The rest of the paper is structured as follows. In Section 2, it shown that hydrodynamics equations for a thin spherical layer are satisfied only by the APV stream function and so APV is the only one relevant elementary singular vortex object on a sphere. In Section 3, dynamics equations of N interacting APVs are given. In Section 4, a non-linear theory of stability of a steady solution for the equations is considered for the case of N=2 APVs to estimate cyclone-anticyclone ACA-like pairs stability.

## 2. REVISION OF THE PROBLEM OF ELEMENTARY SINGULAR VORTEX OBJECT ON A SPHERE

The absolute vorticity conservation equation, which is used in the next Chapter, is derived in [22] (see p.698) on the base of three-dimensional hydrodynamic equations of an ideal incompressible fluid in a thin layer on a rotating globe. In the spherical coordinate system $(r, \theta, \varphi)$, rigidly connected to the globe and with the origin in the center of the globe, and on the assumption that only the radial component of the velocity field is equal to zero $V_r = 0; V_\theta \neq 0; V_\varphi \neq 0$ that equations are represented in the form:

$$\frac{V_\theta^2 + \left(V_\varphi + \Omega r \sin \theta\right)^2}{r} = \frac{1}{\rho_0} \frac{\partial p}{\partial r} \tag{2.1}$$

$$\frac{\partial V_\theta}{\partial t} + \frac{V_\theta}{r} \frac{\partial V_\theta}{\partial \theta} + \frac{\left(V_\varphi + \Omega r \sin \theta\right)}{r \sin \theta} \frac{\partial V_\theta}{\partial \varphi} - \frac{\left(V_\varphi + \Omega r \sin \theta\right)^2 \cot \theta}{r} = -\frac{1}{\rho_0 r} \frac{\partial p}{\partial \theta} \tag{2.2}$$

$$\frac{\partial V_\varphi}{\partial t} + \frac{V_\theta}{r} \frac{\partial \left(V_\varphi + \Omega r \sin \theta\right)}{\partial \theta} + \frac{\left(V_\varphi + \Omega r \sin \theta\right)}{r \sin \theta} \frac{\partial V_\varphi}{\partial \varphi} + \frac{V_\theta \left(V_\varphi + \Omega r \sin \theta\right) \cot \theta}{r} = -\frac{1}{\rho_0 r \sin \theta} \frac{\partial p}{\partial \varphi} \tag{2.3}$$

In (2.1)-(2.3) $\rho_0 = const$ is the density, $\Omega = const$ is the constant speed of rotation of the globe or sphere, and $p$ is the pressure.

It is noted in [22] (see p.697) that the presence of balance equation (2.1) is also important for the realization of two-dimensional fluid motion in thin spherical layer due to non-zero value of radial component of pressure gradient.

Let us use the stream function $\psi$ of the two-dimensional flow in thin spherical layer:

$$V_\varphi = -\frac{1}{r}\frac{\partial \psi}{\partial \theta};$$
$$V_\theta = \frac{1}{r\sin\theta}\frac{\partial \psi}{\partial \varphi} \qquad (2.4)$$

A unit PV on a sphere together with compensating uniform regular vorticity field is considered by, e.g. Bogomolov[4]; Kimura and Okamoto[5]; Dritschel and Boatto[7] (see also [9]), as an elementary singular PV object with the stream function (for simplicity, let us consider axially symmetric case when all terms with derivatives $\partial/\partial\varphi$ are zero):

$$\psi_1 = \frac{\Gamma_1}{4\pi}\ln(\frac{1}{1-\cos\theta}) \qquad (2.5)$$

The stream function (2.5) corresponds to the distribution of the vortex field composed of the sum of a point vortex located at the pole of the sphere with circulation $\Gamma_1$ and uniformly distributed vorticity field with the density having the opposite sign and value $-\Gamma_1/4\pi$. This gives zero integral vorticity on the sphere in accordance with the Kelvin theorem.

We show below (see also [23–25]) that (2.5) does not meet (2.1), (2.2), whereas the stream function of APV satisfies them. The stream function for APV placed at the sphere poles is represented in the form[1,3,4,17]:

$$\psi = \frac{\Gamma_1}{2\pi} \ln(\frac{1+\cos\theta}{1-\cos\theta}) \ . \tag{2.6}$$

Let us consider the steady mode case in (2.1)-(2.3) when, for simplicity, $V_\theta = 0; V_\varphi \neq 0; \partial V_\varphi / \partial t = \partial V_\varphi / \partial \varphi = \partial p / \partial \varphi = 0; \Omega = 0$ (equation (2.3) in this case is valid for any $V_\varphi$ and $p$ ). Then, (2.1), (2.2) yield:

$$\frac{V_\varphi^2}{r} = \frac{1}{\rho_0} \frac{\partial p}{\partial r} , \tag{2.7}$$

$$\frac{V_\varphi^2}{r} \cot\theta = \frac{1}{r\rho_0} \frac{\partial p}{\partial \theta} . \tag{2.8}$$

If (2.7), (2.8) are valid (and the Schwarz theorem is invoked, i.e., $\frac{\partial^2 p}{\partial\theta\partial r} = \frac{\partial^2 p}{\partial r\partial\theta}$ ), differentiation of (2.7) and (2.8) with respect to $\theta$ and $r$ , respectively, gives:

$$\frac{1}{r} \frac{\partial V_\varphi}{\partial \theta} = \frac{\partial V_\varphi}{\partial r} \cot\theta \ . \tag{2.9}$$

Using (2.6), we obtain the solution $V_\varphi = \frac{\Gamma_1}{r\pi \sin\theta}$ , satisfying (2.9) for any $r; \theta$ identically. At the same time, the stream function (2.5) yields:

$$V_\varphi = \frac{\Gamma_1}{4r\pi} \cot(\frac{\theta}{2}) \ . \tag{2.10}$$

Plugging (2.10) into (2.9), we obtain that the velocity field (2.10) does not satisfy (2.9) identically, and, thus, the velocity field (2.10) and corresponding stream function (2.5) used in [4,5,7] does not satisfy the hydrodynamic equations (2.7) and (2.8)

Thus, the stream function (2.5) used in [4,5,7] cannot describe any hydrodynamics flow, whereas the APV stream function of the form (2.6) meets hydrodynamics equations (2.1)-(2.3).

In the case when rotation of a sphere with angular velocity $\Omega$ is taken into account in (2.7) and (2.8), one needs to replace $V_\varphi \rightarrow V_\varphi + \Omega r \sin\theta$, and, accordingly, use the stream functions (2.5) and (2.6) with an additional term $\Psi_0 = -\Omega r^2 \cos\theta$. Again, only (2.6) modified by $\Psi_0$ still satisfies the hydrodynamics equations (2.7), (2.8), modified as follows to take into account the rotation of the sphere:

$$\frac{\left(V_\varphi + \Omega r \sin\theta\right)^2}{r} = \frac{1}{\rho_0}\frac{\partial p}{\partial r};$$
$$\frac{\left(V_\varphi + \Omega r \sin\theta\right)^2 \cot\theta}{r} = \frac{1}{\rho_0 r}\frac{\partial p}{\partial \theta} \tag{2.11}$$

The sum of (2.5) and $\Psi_0$ does not satisfy (2.11). On the other hand, the system of equations (2.11) is exactly satisfied by the APV stream function on a rotating sphere:

$$\psi = -\Omega r^2 \cos\theta + \frac{\Gamma_1}{2\pi}\ln\frac{1+\cos\theta}{1-\cos\theta} \tag{2.12}$$

The stream function (2.12) is used in the following paragraph (see (3.3) for the case of a system of an arbitrary number of APVs. When the approximation of a thin layer on a rotating globe $r - R << R$ ($R$ - radius of the Earth) is used the replacement $r \rightarrow R$ is also used for representation of the stream function in (3.3) on the base of exact solution (2.12).

## 3. APV DYNAMICS ON A ROTATING SPHERE

The dynamic interaction of APVs is considered based on an exact weak solution of the absolute vorticity conservation equation on a rotating sphere which for the case of constant thin of spherical layer is [21,22]:

$$\frac{\partial \omega}{\partial t} + \frac{V_\theta}{R}\frac{\partial \omega}{\partial \theta} + \frac{V_\varphi}{R\sin\theta}\frac{\partial \omega}{\partial \varphi} = 0 \qquad (3.1)$$

where $\omega = \omega_r + 2\Omega\cos\theta$; $\Omega$ is the angular velocity of the sphere rotation (for the Earth, $\Omega \approx 7.3\times 10^{-5}\,\mathrm{sec}^{-1}$), $\theta$ is the co-latitude; $\varphi$ is the longitude;

$V_\theta = R\dfrac{d\theta}{dt}, V_\varphi = R\sin\theta\dfrac{d\varphi}{dt}$; $\omega_r = \dfrac{1}{R\sin\theta}\left(\dfrac{\partial V_\varphi \sin\theta}{\partial \theta} - \dfrac{\partial V_\theta}{\partial \varphi}\right) = -\Delta\psi$ is the radial component of the

local vortex field on the sphere, $\Delta$ is the Beltrami- Laplace operator; $\psi$ is the stream function for which $V_\varphi = -\dfrac{1}{R}\dfrac{\partial \psi}{\partial \theta}, V_\theta = \dfrac{1}{R\sin\theta}\dfrac{\partial \psi}{\partial \varphi}$ in (3.1) (see (2.4) when $r \to R$). Equation (3.1) describes the two-dimensional motion of a thin spherical layer of fluid in the so called rigid cap approximation [21].

As in [21], we use (3.1) as an equation for an unknown stream function, representing the absolute vorticity as $\omega = -\Delta\psi + 2\Omega\cos\theta$, when in (3.1) we will also replace the components of the velocity field with their representation via (2.4).

According to (3.1), each Lagrangian particle preserves the value of absolute vorticity in a thin

layer of liquid on a rotating sphere. Therefore, as in [21,23–25], we use the absolute vortex field in the

form of APVs system:

$$\omega = -\Delta\psi + 2\Omega\cos\theta = \hat{L}(\delta);$$
$$\hat{L}(\delta) = \frac{1}{R^2}\sum_{i=1}^{N}\frac{\Gamma_i}{\sin\theta_i}\left(\delta(\theta - \theta_i)\delta(\varphi - \varphi_i) - \delta(\theta + \theta_i - \pi)\delta(\varphi - \varphi_i - \pi)\right) \qquad (3.2)$$

In (3.2) $\delta$ is the Dirac delta-function. To determine the type of stream function that satisfies

(3.2), we present it as a superposition $\psi = \psi_0 + \psi_{APV}$. In this case, we use the function

$\psi_0 = -\Omega R^2\cos\theta$ for which the value of absolute vorticity is zero $-\Delta\psi_0 + 2\Omega\cos\theta = 0$.

Therefore, an equation $-\Delta \psi_{APV} = \hat{L}(\delta)$ is obtained from (3.2), the known [3,5,17,19] solution for which for the stream function of a system of $N$ APVs has the form

$$\psi_{APV} = \frac{1}{2\pi} \sum_{i=1}^{N} \Gamma_i \ln \frac{1+\cos u_i}{1-\cos u_i}; \cos u_i = \cos\theta\cos\theta_i + \sin\theta\sin\theta_i\cos(\varphi-\varphi_i); \Gamma_i = const.$$

As a result, we obtain an exact solution of equation (3.2) for the stream function in the form:

$$\psi = -\Omega R^2 \cos\theta + \frac{1}{2\pi} \sum_{i=1}^{N} \Gamma_i \ln \frac{1+\cos u_i}{1-\cos u_i} \qquad (3.3)$$

Note that in the case of $N=1$ the stream function (3.3) exactly coincides with the stream function (2.12) for a single APV located at the poles of the sphere.

In (3.3) $\theta_i, \varphi_i$ are the spherical coordinates of the APV may be time-dependent.

The stream function $\psi = \psi_0 = -\Omega R^2 \cos\theta$ is the solution of (3.1) which describes the solid-state rotation of fluid in the direction opposite (like anticyclone) to the direction of the sphere rotation because $V_{\varphi 0} = \dfrac{d\varphi}{dt} = -\dfrac{1}{R}\dfrac{\partial\psi_0}{\partial\theta} = -\Omega R\sin\theta < 0$ for $\Omega > 0$. This global solid-state rotation of fluid layer has the sign different from the direction of the background solid-state rotation introduced in [20] without any relation to absolute vorticity conservation equation (3.1).

The Kelvin theorem requirement that the integral of vorticity on the sphere is equal to zero ($\int_0^{2\pi} d\varphi \int_0^{\pi} d\theta \sin\theta\, \omega = 0$) holds identically for all values $\Gamma_i, i = 1...N$, in the form (3.2).

The vorticity field (3.2) and stream function (3.3) may be used for obtaining an exact weak solution (in the sense of generalized functions) to (3.1). A weak solution to (3.1) is obtained by substituting in (3.1) the vortex field (3.2) and corresponding steam function (3.3). Multiplying the resulting expression by an arbitrary smooth function and integrating over angular spherical coordinates, the following system of ordinary differential equations for APV coordinates is obtained, representing an exact weak solution for (3.1). As a result, the functions $\theta_i(t),\, \varphi_i(t)$ are defined as the solutions to the following *2N*-dimensional Hamiltonian system of ordinary differential equations [23–25]:

$$\frac{d\theta_i}{dt} = \dot{\theta}_i = -\frac{1}{\pi R^2} \sum_{\substack{k=1 \\ k \neq i}}^{N} \frac{\Gamma_k \sin\theta_k \sin(\varphi_i - \varphi_k)}{1 - \cos^2 u_{ik}};$$

$$\cos u_{ik} = \cos\theta_i \cos\theta_k + \sin\theta_i \sin\theta_k \cos(\varphi_i - \varphi_k)$$

$$\frac{d\varphi_i}{dt} = \dot{\varphi}_i = -\Omega - \frac{1}{\pi R^2} \sum_{\substack{k=1 \\ k \neq i}}^{N} \frac{\Gamma_k \left( \cot\theta_i \sin\theta_k \cos(\varphi_i - \varphi_k) - \cos\theta_k \right)}{1 - \cos^2 u_{ik}} \qquad (3.5)$$

The system (3.5) for $\Omega = 0$ with accuracy up to a multiplier (set to $\pi$) coincides with the corresponding system derived in [17] for N pairs of APVs.

In the case under consideration, the energy dissipation and pumping in the system are treated as insignificant or balancing each other. The system (3.5) conserves integral invariants of kinetic energy $\overline{E}$, angular momentum $\overline{\boldsymbol{M}}$, and impulse, $\overline{\boldsymbol{P}}$, where the line above the variables denotes the averaging process for the respective variable over the surface of a sphere. The noted values (per unit of mass) in a rotating coordinate system have the following form: $E = \frac{1}{2}\boldsymbol{V}^2$, $\boldsymbol{M} = [\boldsymbol{r} \times \boldsymbol{V}]$, $\boldsymbol{P} = \boldsymbol{V}$, where $\boldsymbol{V} = \dfrac{d\boldsymbol{r}}{dt}$ and $\boldsymbol{r}$ is the position vector in a Cartesian coordinate system $(x, y, z)$ whose origin is at the center of a sphere.

From the definition, $\overline{E} = \int\limits_0^{2\pi} d\varphi \int\limits_0^{\pi} d\theta \sin\theta E = \frac{1}{2} \int\limits_0^{2\pi} d\varphi \int\limits_0^{\pi} d\theta \sin\theta \,\psi\, \omega_r$, it follows that $\overline{\boldsymbol{P}} = 0$, and also:

$$\overline{E} = \frac{1}{8\pi} \sum_{i=1}^{N} \sum_{\substack{k=1 \\ k \neq i}}^{N} \frac{\Gamma_i \Gamma_k}{R^2} \ln \frac{1 + \cos u_{ik}}{1 - \cos u_{ik}} - \Omega \sum_{i=1}^{N} \Gamma_i \cos \theta_i + const \qquad (3.6)$$

$$\overline{M}_z = 2 \sum_{i=1}^{N} \Gamma_i \cos \theta_i \qquad (3.7)$$

$$\overline{M}_x = 2 \sum_{i=1}^{N} \Gamma_i \cos \varphi_i \sin \theta_i ; \overline{M}_y = 2 \sum_{i=1}^{N} \Gamma_i \sin \varphi_i \sin \theta_i \qquad (3.8)$$

The values of $\theta_i$ and $\varphi_i, i = \overline{1, N}$, in (3.6)-(3.8) are functions of time derived from the dynamic equations (3.5) for the corresponding initial conditions. For $\Omega = 0$, all four values in (3.6)-(3.8) are invariants of the system (3.5) and exactly coincide with the invariants of the dynamic system in[3, 4, 17]. For $\Omega \neq 0$ the values of (3.8) are already not invariant, but values $\overline{E}$ and $\overline{M}_z$ are still invariants of (3.5).

## 4. STEADY VORTEX MODES (*N*=2) AND THEIR STABILITY

Let us consider steady modes corresponding to the equilibrium for $N = 2$ APVs in (3.5). If $\dot{\theta}_1 = \dot{\theta}_2 = 0$, $\theta_1 = \theta_{10} = const; \theta_2 = \theta_{20} = const$ absolute equilibrium is possible when $\varphi_1 = \varphi_2 = const$, while $\varphi_1 - \varphi_2 = \pi$ corresponds to relative equilibrium when APVs may move along there constant latitudes with constant velocity as in[21].

From (3.5) it is possible to obtain the condition of relative equilibrium when $\frac{d(\varphi_1 - \varphi_2)}{dt} = 0$:

$$\frac{\Gamma_1}{\Gamma_2} = -\frac{\sin \theta_{20}}{\sin \theta_{10}} \qquad (4.1)$$

Condition (4.1) exactly coincides with the relative equilibrium condition obtained in[21] for a system of two vortices (see (15) in[21] since the coefficients $A_i$, used in[21] to denote the values of the circulation $\Gamma_i$ of point vortices are related to the values by the relation $A_i = \Gamma_i \sin \theta_i$).

An additional condition for the absence of absolute motion uses the equality $\dfrac{d(\varphi_1 + \varphi_2)}{dt} = 0$, and has the form:

$$\Gamma_2 = -\Omega \pi R^2 \sin \theta_{10} \sin(\theta_{20} - \theta_{10}) < 0; \Omega > 0, \theta_{20} > \theta_{10}. \qquad (4.2)$$

From (4.1), it follows that $\gamma_1 = \Gamma_1 / \Gamma_2 < 0$. It means that we consider a steady mode for two APVs having opposite circulation directions and placed on the same meridian (since the condition (4.1) is obtained when $\boldsymbol{\varphi_{10} = \varphi_{20}}$). Also, from (4.1), it follows that $|\Gamma_2| < |\Gamma_1|$ and, thus, APV with $\theta = \theta_{20}$ has less intensity than APV with $\theta = \theta_{01}$, if, by definition, we assume $\theta_{20} > \theta_{10}$ in (4.1) and (4.2). Moreover from (4.2) we can determine that APV with the coordinate $\boldsymbol{\theta_{20}}$ and intensity $\boldsymbol{\Gamma_2}$ is indeed anticyclone because an anticyclone having the direction of rotation opposite to the direction of the rotation of the sphere.

Indeed, for the ACA over the North Atlantic, the meridional coordinates of the Icelandic cyclone and the Azores anticyclone are close each to other. This corresponds to the case $\varphi_1 = \varphi_2 = const$ under which the conditions (4.1) and (4.2) are obtained. According to (4.1), ACA over North Atlantic have opposite signs and different values of the vortex circulation. As in (4.1), in reality, the intensity of the Icelandic cyclone usually exceeds the intensity of the Azores anticyclone.

Consider now stability of the obtained[23-25] steady mode (4.1), (4.2), modeling steady regime of a cyclonic-anticyclonic ACA pair. We use for this the system (3.5) for the case N=2 and $\Gamma_0 = 0$. Let us introduce a disturbance of the steady mode (4.1), (4.2) as

$$x = \theta_1(t) - \theta_{10}; z = \theta_2(t) - \theta_{20}; y = \varphi_1(t) - \varphi_2(t). \qquad (4.3)$$

Then, from (3.5), taking into account (4.1), (4.2), the system of equations for the disturbances (4.3) is as follows:

$$\frac{dx}{d\tau} = \sin\theta_{10}\sin(\theta_{20}+z)\sin y,$$

$$\frac{dz}{d\tau} = \sin\theta_{20}\sin(\theta_{10}+x)\sin y,$$ (4.4)

$$\frac{dy}{d\tau} = -\sin\theta_{10}\cos(\theta_{20}+z) - \sin\theta_{20}\cos(\theta_{10}+x) + W\cos y,$$

$$W = \sin\theta_{10}\cot(\theta_{10}+x)\sin(\theta_{20}+z) + \sin\theta_{20}\cot(\theta_{20}+z)\sin(\theta_{10}+x)$$

$$\tau = t\Omega\frac{\sin(\theta_{20}-\theta_{10})}{1-U_0^2};$$

$$U_0 \equiv U(\tau=0) = \cos(\theta_{10}+x(0))\cos(\theta_{20}+z(0)) + \sin(\theta_{10}+x(0))\sin(\theta_{20}+z(0))\cos y(0)$$

The system (4.4) describes non-linear evolution of the disturbances and, taking into account energy invariant (3.6), it yields:

$$U = \cos u_{12} = \cos(\theta_{10}+x)\cos(\theta_{20}+z) + \sin(\theta_{10}+x)\sin(\theta_{20}+z)\cos y = U_0 = const. \quad (4.5)$$

In (4.4), intensities $\Gamma_1, \Gamma_2$ of the vortices are defined under conditions (4.1), (4.2). The value $\theta_{10}$ corresponds to the latitudinal steady position of a cyclone (like Icelandic cyclone) with disturbance $x(t)$, and the value $\theta_{20}$ defines the steady position of an anticyclone (like Azores anticyclone), disturbance of which is denoted as $z(t)$. Difference of the longitudes of cyclone and anticyclone is characterized by $y(t)$ in (4.4).

The system (4.4), in addition to (4.5), has also an invariant (3.7), which provides the following relation between arbitrary by amplitude disturbances, $z(t)$ and $x(t)$

$$C(x) = \cos(\theta_{20}+z(t)) = M_0 + \frac{\sin\theta_{20}}{\sin\theta_{10}}\cos(\theta_{10}+x(t)),$$

$$M_0 = \cos(\theta_{20}+z(0)) - \frac{\sin\theta_{20}}{\sin\theta_{10}}\cos(\theta_{10}+x(0)) = const. \quad (4.6)$$

In the limit of small linear disturbances, (4.6) yields: $z(t) - z(0) = x(t) - x(0)$.

From (4.4) and (4.1), (4.2), in the limit of small disturbances, $x << 1, y << 1$ we get

$$\frac{dx}{d\tau} = y \sin \theta_{10} \sin \theta_{20},$$

$$\frac{dy}{d\tau} = -x \frac{\sin^2(\theta_{20} - \theta_{10})}{\sin \theta_{10} \sin \theta_{20}} - C, \qquad (4.7)$$

$$C = (x(0) - z(0)) \cot \theta_{20} \sin(\theta_{20} - \theta_{10}).$$

From (4.7), we get $\frac{d^2 x}{d\tau^2} + x \sin^2(\theta_{20} - \theta_{10}) = -C_1 \sin^2(\theta_{20} - \theta_{10}); C_1 = \frac{(x(0) - z(0)) \cos \theta_{20} \sin \theta_{10}}{\sin(\theta_{20} - \theta_{10})}$,

solution of which is $x(\tau) = -C_1 + (x(0) + C_1) \cos \left[ \tau \sin(\theta_{20} - \theta_{10}) \right]$. For this solution frequency of oscillations in the dimensional form is as follows $\Omega_0 = \Omega \frac{\sin^2(\theta_{20} - \theta_{10})}{1 - U_0^2}$. .

Thus, it follows that the steady mode is a center type equilibrium point, in the proximity of which oscillations have frequency that depends on the latitude of the modeled vortex and also on the initial disturbance of their mutual positions on the longitude $y(0)$. In particular, if $y(0) = 0$ then it follows that small oscillations about the equilibrium have the frequency of the sphere rotation, $\Omega_0 = \Omega$.

Generally, for not small finite-amplitude disturbances, the unknown disturbance, y(t), can be expressed via x(t), if using invariants (4.5) and (3.7),

As a result, for the finite-amplitude initial disturbance, from (4.4)-(4.6), we get just one equation for the disturbance $\boldsymbol{x(t)}$:

$$\frac{du}{d\tau} = -\sin\theta_{10}\sqrt{-\hat{c}u^2 + \tilde{b}u - \tilde{a}};$$
$$u = \cos(\theta_{10} + x(\tau));$$
$$\tilde{c} = (S_0 - U_0)^2 + 1 - U_0^2; S_0 = \frac{\sin\theta_{20}}{\sin\theta_{10}}; \tilde{b} = 2|M_0|(S_0 - U_0);$$
$$\tilde{a} = M_0^2 + U_0^2 - 1.$$

$$(4.8)$$

The exact common solution of (4.8) is obtained in the form:

$$u(\tau) = \cos(\theta_{10} + x(\tau)) = A - B\sin(\tau\Omega_1 + \Phi);$$
$$A = \frac{\tilde{b}}{2\tilde{c}}; B = \sqrt{\Delta}/2\tilde{c}; \Omega_1 = \sqrt{\Delta}\sin\theta_{10};$$
$$\Phi = \frac{\tilde{b}}{\sqrt{\Delta}} - 2\frac{\tilde{c}}{\sqrt{\Delta}}\cos(\theta_{10} + x(0)); \Delta = \tilde{b}^2 - 4\tilde{a}\tilde{c},$$

$$(4.9)$$

It follows from (4.9), that the steady mode modeling ACA has vortex circulation defined in (4.1), (4.2), and is stable also with respect to finite-amplitude disturbances.

Thus, on the base of the derived steady solution (4.1), (4.2) and analysis of its stability, we have proposed a hydrodynamic model explaining observed stability for cyclonic-anticyclonic vortices like ACA.

Despite the noted stability of the dynamics of the perturbed vortex system, the relative magnitude of change of the vortices location during their oscillations near the stationary state substantially depends on the form of the initial perturbations and their amplitude. The sensitivity to and variability of the vortex system dependence on the nature of the initial perturbations is already visible from the linear solutions according which, the maximum deviation from the equilibrium state of the perturbed vortex system depends on the initial perturbations of the cyclone vortex $x(0)$ and anticyclone vortex $z(0)$ :

$$x_{max} = -x(0) - 2\frac{(x(0) - z(0))\cos\theta_{20}\sin\theta_{10}}{\sin(\theta_{20} - \theta_{10})},$$

$$(4.10)$$

$$z_{max} = z(0) - 2x(0) - 2\frac{(x(0) - z(0))\cos\theta_{20}\sin\theta_{10}}{\sin(\theta_{20} - \theta_{10})}$$

In this case, the maximum displacement relative to the equilibrium position for the anticyclonic vortex noticeably differs from the displacement amplitude of the cyclonic vortex for any small initial perturbation. From (4.10) it follows that only in the case of equality of the initial perturbations $x(0) = z(0)$, the maximum deviation of the vortices from the stationary position is equal to the amplitude of the initial perturbation $x_{max} = z_{max} = -x(0)$.

With a perturbation of the position of only the cyclonic vortex, the displacement of the center of the anticyclonic vortex can significantly (up to three times) exceed the value of the initial perturbation. This effect becomes most noticeable with the perturbation amplitude increasing. For example, in the case corresponding to the average position of the Icelandic cyclone and the Azores anticyclone, the exact nonlinear solution (4.9) gives:

$$u(\tau) = 0.917 - 0.067\sin(\tau\Omega_1 + \Phi);$$
$$C(\tau) = 0.708 - 0.13\sin(\tau\Omega_1 + \Phi).$$

In this case, the maximum value corresponds to a shift of the cyclonic vortex in the north direction by an amount that is almost twice the amplitude of its initial shift in the south direction. In this case, the maximum value corresponds to the possibility of displacement of the anticyclonic vortex (from its equilibrium position at $\boldsymbol{\theta_{20} = 55^0}$) in the north direction by an amount that is already three times higher than the amplitude of the initial perturbation. In the case of $\boldsymbol{x(0) = z(0) = 7^0; y(0) = 0}$, the deviation of the vortices from their stationary state have values not exceeding the amplitude of the initial perturbation, which follows from the linear solution (4.10).

The conclusion obtained about the nonlinear stability of the stationary regime of a vortex pair on a rotating sphere does not exclude the possibility of a significant change in the position of the center of the anticyclonic vortex for certain initial perturbations of the cyclonic vortex (IC) with its shift to the south (by 5 °), causing a significant shift of the anticyclonic vortex (AZ) to the north (by 21 °). At the same time, with the same magnitude and direction of the initial displacement of the anticyclonic vortex (AZ) by 5 degrees to the South, the subsequent evolution of the vortices does not lead to their deviation from the equilibrium position by more than 5 degrees (see Fig.1). Fig.1 shows the results of integration of both the initial nonlinear equations (4.4) and linear equations (4.7) obtained from (4.4) when linearizing near the equilibrium state (4.1), (4.2). It can be seen that for the considered amplitude of initial perturbations of 5 degrees, the difference between solutions of linear and nonlinear systems is insignificant.

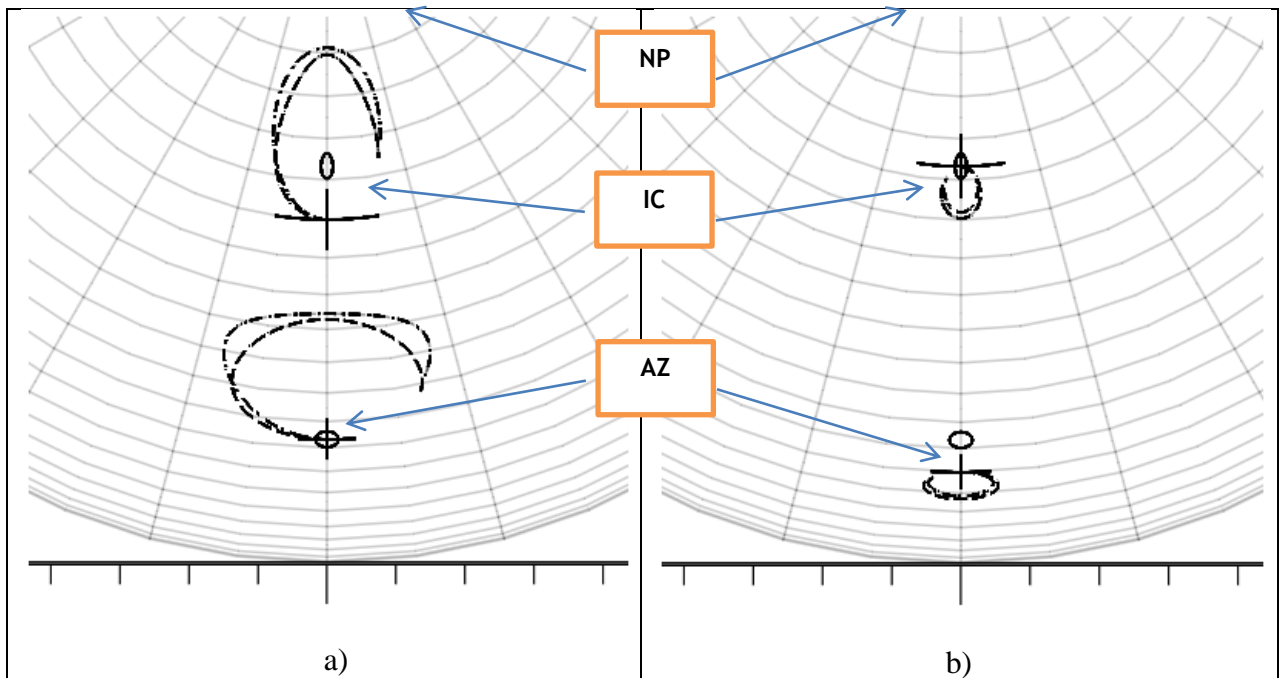

Fig. 1. Plots of periodic trajectories of Icelandic (IC, about equilibrium $\theta_{10} = 25^0$ shown by a small solid ellipse) and Azores (AZ, about equilibrium $\theta_{20} = 55^0$ shown by a small solid ellipse) vortices according to (4.4) and its linearization (4.7) for various initial conditions (shown by crosses) in the Northern hemisphere. Simulation time is 0..10, less than a period. a) x(0)=5$^0$, z(0)=y(0)=0; solutions of (4.7), dash, mainly inner lines, and solutions of (4.4), dashdot, mainly outer lines; b) z(0)=5$^0$, x(0)=y(0)=0, solutions of (4.7) are outer lines, and solutions of (4.6) are inner lines The North Pole (NP) is the point of the meridians (radial lines with stride 15$^0$) intersection (not visible). Concentric lines represent latitudes with stride 3.75$^0$.

Thus, significant weather and climatic anomalies can be associated only with initial displacements of the Icelandic cyclone in the south direction (see Fig.1a). It is interesting to consider the obtained above estimations in comparison with observations [27–31].

According to a long-term observations (see Appendix C in [31]) the cyclonic-anticyclonic vortex pair of the ACAs over North Atlantic, usually only slightly deviates from the equilibrium annual-mean coordinates that corresponds to the case illustrated by Fig. 1b. However, for some years, significant deviations are observed as shown in Fig. 1a. The mechanism for the formation of such dynamics with significant difference in behavior, as manifested for atmospheric centers of action in Fig. 1a and Fig. 1b, requires clarification.

Until now, we have not taken into account the permanently existing polar vortices which may affect substantially stability of the equilibrium state of the above mentioned paired ACAs even in a linear approximation [31].

In the next Section 5, it is shown that accounting for the finite intensity of the polar vortices leads to the opportunity of the stationary state linear instability. The instability is likely to be a natural mechanism determining the conditions for emerging of the observed substantial deviations from the equilibrium shown in Fig. 1a. Really, for any type small initial disturbance, in the case of exponential instability, it is always possible realization of the deviation from the equilibrium of the kind as in Fig.1a, resulting in significant deviation.

### 5. Polar vortex and stability observation data.

Let us consider the case when a cyclonic-anticyclonic pair of APV interacts with an APV located on the poles with the stream-function $\psi_P = \dfrac{\Gamma_0}{2\pi} \ln\left(\dfrac{1+\cos\theta}{1-\cos\theta}\right)$ to be added in the right-hand side of (3.3) [31,25]. Then, it is necessary only make the following substitutions

$$\dot{\varphi}_i \rightarrow \dot{\varphi}_i + \Gamma_0 / \pi R^2 \sin^2 \theta_i ;$$
$$\overline{M}_z \rightarrow \overline{M}_z + 4\Gamma_0 ;$$
$$\overline{E} \rightarrow \overline{E} - 2\Omega\Gamma_0 + \frac{\Gamma_0}{2\pi R^2} \sum_{k=1}^{N} \Gamma_k \ln\left(\frac{1+\cos\theta_k}{1-\cos\theta_k}\right) \qquad (5.1)$$

in (3.5)-(3.7). Taking into account (5.1), instead of equilibrium conditions (4.1) and (4.2), the new conditions obtained for the realization of the stationary state of the cyclonic-anticyclonic pair of APV on the rotating sphere with a polar APV having non-zero circulation $\Gamma_0$ :

$$\frac{\Gamma_0}{\Gamma_2} = \left(\frac{\Gamma_1}{\Gamma_2} + \frac{\sin\theta_{20}}{\sin\theta_{10}}\right) \frac{\sin^2\theta_{10}}{\left(\sin^2\theta_{20} - \sin^2\theta_{10}\right)\sin(\theta_{20} - \theta_{10})} \ , \qquad (5.2)$$

$$\frac{\Gamma_0}{\Gamma_2} = \frac{\sin\theta_{10}\sin\theta_{20}}{\left(\sin^2\theta_{10} + \sin^2\theta_{20}\right)}\left[\frac{\sin\theta_{20}}{\sin(\theta_{20} - \theta_{10})} - \frac{\Gamma_1}{\Gamma_2}\frac{\sin\theta_{10}}{\sin(\theta_{20} - \theta_{10})} + \frac{2\pi R^2 \Omega \sin\theta_{10}\sin\theta_{20}}{\Gamma_2}\right]. \qquad (5.3)$$

Then, in the case of $\Gamma_0 = 0$, equations (5.2), (5.3) exactly coincide with the equilibrium conditions (4.1), (4.2).

Stability conditions for the stationary state (5.2), (5.3) in the limit of extremely small disturbances are as follows [25,31]:

$$D = A\gamma_1^2 + 2B\gamma_1 + C < 0; \gamma_1 = \frac{\Gamma_1}{\Gamma_2};$$
$$A = \sin^3\theta_{10}\left[\sin(2\theta_{20} - \theta_{10}) + \frac{2\sin^2\theta_{10}\cos\theta_{20}\sin(\theta_{20} - \theta_{10})}{\sin^2\theta_{20} - \sin^2\theta_{10}}\right];$$
$$B = \sin\theta_{10}\sin\theta_{20}\left(\sin^2\theta_{20} + \sin^2\theta_{10}\right); \qquad (5.4)$$
$$C = \sin^3\theta_{20}\left[\sin(2\theta_{10} - \theta_{20}) + \frac{2\sin^2\theta_{20}\cos\theta_{10}\sin(\theta_{20} - \theta_{10})}{\sin^2\theta_{20} - \sin^2\theta_{10}}\right]$$

In particular, for the equilibrium coordinates of the vortex pair of APV $\theta_{20} = 55^0 ; \theta_{10} = 25^0$, the condition (5.4) yields the following inequality:

$$1.36 < \left| \frac{\Gamma_1}{\Gamma_2} \right| < 5.11 \qquad (5.5)$$

Then the intensity of the polar vortex corresponding to the stationary state (5.2), (5.3) is $\Gamma_0 / \Gamma_2 = 0.725 \left( 1.94 - \left| \Gamma_1 / \Gamma_2 \right| \right) ; \Gamma_1 > 0 ; \Gamma_2 < 0$. For $\left| \Gamma_1 / \Gamma_2 \right| > 1.94$, it corresponds to the cyclonic circulation orientation of the polar vortex. Vice versa, anticyclonic circulation of the polar vortex corresponding to observations [31] satisfies $\left| \gamma_1 \right| = \left| \Gamma_1 / \Gamma_2 \right| < \sin \theta_{20} / \sin \theta_{10}$ following from (5.2).

If the condition (5.4) is violated, i.e. when for $D > 0$, the stationary state (5.2), (5.3) is exponentially unstable and the following relationships hold:

$$x(t) = \theta_1(t) - \theta_{10} = x(0) \exp(t\lambda);$$
$$\lambda = \frac{\left| \Gamma_2 \right|}{\pi R^2} \frac{\sqrt{D}}{\sin^2(\theta_{20} - \theta_{10}) \sin \theta_{20} \sin \theta_{10}} \qquad (5.6)$$

According to (5.5), instability with respect to extremely small disturbances takes place when inequalities (5.5) are violated for $\left| \Gamma_1 / \Gamma_2 \right| > 5.24$ or $\left| \Gamma_1 / \Gamma_2 \right| < 1.35$. For example, if $\left| \Gamma_1 / \Gamma_2 \right| \approx 1.3$ in (5.6), then $\sqrt{D} \approx 0.14$ that allows estimating the character time of the disturbance growth $\lambda^{-1} \approx 14 days$ if accepting that anticyclonic APV intensity is $\left| \Gamma_2 \right| \approx RU ; U \approx 10 m / \sec$ $R \approx 6371 km$.

It means that in the period of about one month even in the case of small disturbances with amplitude about $1^0$ the deviation from the equilibrium up to $5^0$ is possible with manifestation of a dynamic mode like in Fig. 1a.

Taking into account of the effect of a stationary polar vortex on the dynamics of the APV vortex pair on a rotating sphere is important also in view of the opportunity of generalizing the problems considered in [32,33]. It is already for the case of global atmospheric vortices, when it is necessary taking into account spherical geometry of the motion and the sphere rotation.

Results of comparison of the stability condition (5.4) with observations [31] are presented below. An analysis was conducted in [31] for the variability in the mean distance between ACAs by longitude, $\Delta\lambda$ for the Atlantic (Icelandic Low and Azores High) and Pacific (Aleutian Low and Hawaiian High) ACAs as abnormality characteristics of the position and instability of the vortex pairs. In particular, reanalysis data were used for the detection of ACA characteristics similar to [27]. The abnormality (instability) of mutual ACA positioning during the particular season was characterized in [31] by deviation from the long-term mean against the standard deviation. Also the degree of abnormality (instability) in the temperature difference between land and ocean $\Delta T$ in the Northern Hemisphere was estimated with respect to the mean conditions from CRU data for winters (http://www.uea.ac.uk/cru/data). The obtained estimates show the comparability of the dynamic and thermal factors in the formation of stability modes or instability of the ACA mutual positioning on the sphere.

On Fig. 2 and Fig. 3 show stability regions depending on the positions of anticyclonic and cyclonic vortices according to condition (5.4). The horizontal axis on Figs. 2,3 corresponds to the co-latitude $\theta$ for the anticyclonic ACA vortex center, and the vertical axis for that of the cyclonic ACA. Crosses on figures characterize mean values of $\theta$ for the respective vortices of the ACA pair (by the cross position) and their standard deviations (by the cross size in the respective direction). Dark region shows stability obtained according to the condition (5.4). The stationary mode of the vortex pair on Fig. 2 (similar to Icelandic-Azores ACA pair in 1988) falls in the stability region, and on Fig. 3 (similar to Icelandic-Azores ACA pair in 1964) does not fall into it. Shadowed regions on Figs. 2, 3 correspond to the conditions of realization of exactly anticyclonic circulation of the polar vortex in the Northern Hemisphere when the right hand side of (5.2) and (5.3) is positive.

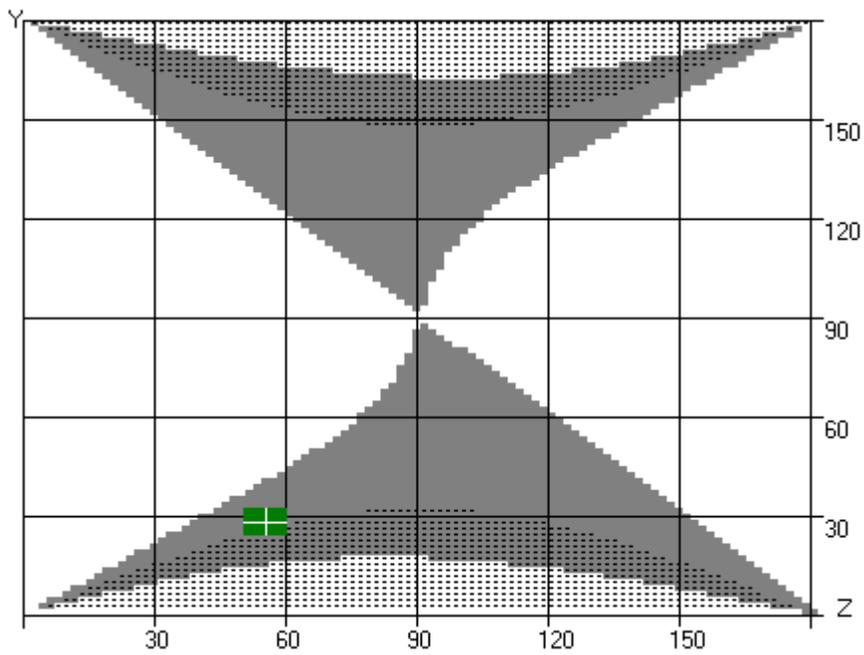

Fig. 2 Stability regions for $\gamma_1 = \dfrac{\Gamma_1}{\Gamma_2} = -1.87$ with co-latitudinal position of vortex pair at

$\theta_{10} = 30^0; \theta_{20} = 57.5^0$ in the stability area.

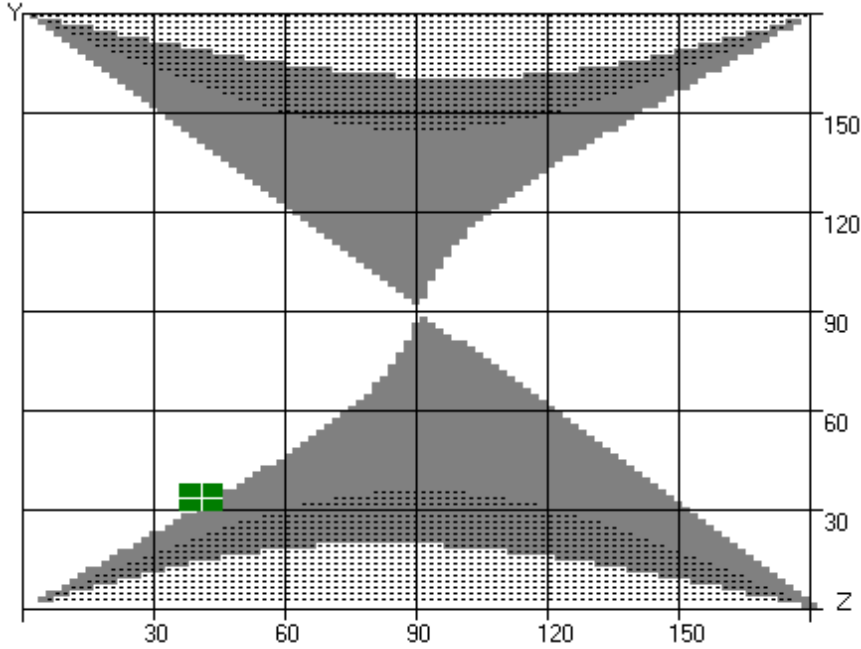

Fig. 3. Stability regions for $\gamma_1 = \dfrac{\Gamma_1}{\Gamma_2} = -1.71$ with co-latitudinal position of vortex pair at

$\theta_{10} = 35^0 ; \theta_{20} = 42.5^0$ in the instability area.

On the whole, the comparison of the theoretical results with long-term observational data[31] indicates their general agreement.

**CONCLUSION**

It is shown that only APV meets the hydrodynamic equations on a sphere. The stream functions used in[2-7] do not meet the hydrodynamic equations (2.1)-(2.3), so it cannot be realized in fluid dynamics. Thus, APV is the only correct elementary vortex object on a sphere.

An exact solution for the APVs dynamics on the rotating sphere is obtained by using stream function of solid-state rotation, corresponding to zero absolute vorticity. On this base a steady solution for two APVs is obtained which gives a possible hydrodynamic mechanism for long-living stable cyclonic-anticyclonic vortex pairs like IC and AA and their dynamics. The possible sensitivity of IC and AA centers of atmospheric action over North Atlantic to the different type of initial disturbances is also explained on the base of linear and nonlinear stability analysis of the steady solution obtained.

The authors thank Anthony R. Lupo for helpful comments.

Data Availability Statements: **"**The data that support the findings of this study are available from the corresponding author upon reasonable request."